\documentclass[preprint,showpacs,preprintnumbers,amsmath,amssymb]{revtex4}

\usepackage{graphicx}
\usepackage{bm}
\graphicspath{{Fig/}}

\begin{document}
\title{Distinct Superconducting Gap on Two Bilayer-Split Fermi Surface Sheets in  $Bi_2Sr_2CaCu_2O_{8+\delta}$ Superconductor}

\author{Ping Ai$^{1,2}$, Qiang Gao$^{1,2}$, Jing Liu$^{1,2}$, Yuxiao Zhang$^{1,2}$, Cong Li$^{1,2}$, Jianwei Huang$^{1,2}$, Chunyao Song$^{1,2}$, Hongtao Yan$^{1,2}$,  Lin Zhao$^{1,2}$, Guodong Liu$^{1,2,5}$,  Genda Gu$^{3}$, Fengfeng Zhang$^{4}$, Feng Yang$^{4}$, Qinjun Peng$^{4}$, Zuyan Xu$^{4}$ and X. J. Zhou$^{1,2,5,6,*}$}

\affiliation{
\\$^{1}$National Lab for Superconductivity, Beijing National Laboratory for Condensed Matter Physics, Institute of Physics,
Chinese Academy of Sciences, Beijing 100190, China
\\$^{2}$University of Chinese Academy of Sciences, Beijing 100049, China.
\\$^{3}$Brookhaven National Laboratory, Condensed Matter Physics and Materials Science Department, New York, NY 11973, USA.
\\$^{4}$Technical Institute of Physics and Chemistry, Chinese Academy of Sciences, Beijing 100190, China.
\\$^{5}$Songshan Lake Materials Laboratory, Dongguan, Guangdong 523808, China.
\\$^{6}$Beijing Academy of Quantum Information Sciences, Beijing 100193, China
\\$^{*}$Corresponding author: XJZhou@iphy.ac.cn
}
\date{\today}


\begin{abstract}
High resolution laser-based angle-resolved photoemission measurements were carried out on an overdoped $\rm Bi_2Sr_2CaCu_2O_{8+\delta}$  superconductor with a T$_c$ of 75 K. Two Fermi surface sheets caused by bilayer splitting are clearly identified with rather different  doping levels: the bonding  sheet corresponds to a doping level of 0.14 which is slightly underdoped while the antibonding sheet has a doping of 0.27 that is heavily overdoped, giving an overall doping level of 0.20 for the sample. Different superconducting gap sizes on the two Fermi surface sheets are revealed for the first time. The superconducting gap on the antibonding Fermi surface sheet follows a standard {\it d}-wave form while it deviates from the standard {\it d}-wave form for the bonding Fermi surface sheet. The maximum gap difference between the two Fermi surface sheets near the antinodal region is $\sim$2 meV. These observations provide important information for studying the relationship between the Fermi surface topology and superconductivity, and the layer-dependent superconductivity in high temperature cuprate superconductors.
\end{abstract}

\maketitle


High temperature superconductivity in cuprates is believed to be realized by introducing an appropriate amount of charge carriers into the CuO$_2$ planes\cite{Damascelli2003ZXShen,CampuzanoReview, PLee2006Review}.  It has been found that both the normal state properties and superconductivity are rather sensitive to the  doping level. In the underdoped region, the normal state deviates significantly from the Fermi liquid behaviors and the superconducting gap is particularly anomalous: it decreases with increasing doping even though  T$_c$ still increases with doping, and the measured momentum-dependent superconducting gap deviates more obviously from the standard {\it d}-wave form with decreasing doping\cite{Damascelli2003ZXShen, CampuzanoReview, GASawatzky2008SHuefner, ZXShen2012IMVIshik}.  In the overdoped region, on the other hand, the normal state properties appear to become close to a Fermi liquid  while the superconducting gap tends to resemble the BCS behaviors: it decreases with increasing doping with a concomitant T$_c$ decrease and its momentum dependence follows a standard {\it d}-wave form\cite{Damascelli2003ZXShen, CampuzanoReview, ZXShen2012IMVIshik, ZXShen2018YHe}. It has been also found that superconductivity of the cuprate superconductors is sensitive to the number of CuO$_2$ planes, n, in one structural unit; T$_c$ increases from single-layer (n=1), to bilayer (n=2) and trilayer (n=3), reaches a maximum for trilayer (n=3), and then drops with the further increase of the number of CuO$_2$ planes\cite{SUchida1990Tokura, MGreven2004HEisaki, YYWang2016WRuan}. Investigations on the doping dependence and CuO$_2$ layer dependence of the electronic structure and superconducting gap symmetry are important for understanding the origin of high temperature superconductivity in cuprate superconductors

Angle-resolved photoemission spectroscopy (ARPES)  measurements on $\rm Bi_2Sr_2CaCu_2O_{8+\delta}$ (Bi2212) have provided major results on the doping and temperature evolutions of electronic structure, many-body effects, peudogap and superconducting gap about cuprate superconductors\cite{Damascelli2003ZXShen, CampuzanoReview, XJZhouReview, ZXShen2012IMVIshik, ZXShen2014MHashimoto,ZXShen2018YHe} because Bi2212 is easy to cleave to get smooth sample surface for ARPES measurements and can cover a relatively wide range of doping levels. In fact, Bi2212 consists of two CuO$_2$ planes in one structural unit separated by calcium (Ca). The interaction between these two structurally equivalent CuO$_2$ planes gives rise to two Fermi surface sheets, bonding and antibonding, that encompass different Fermi surface areas and thus  different doping concentrations\cite{FPaulsen1995OKAndersenNew, JWheatley1996TXiang, JCooper1998TXiang, ABansil1999PRL, ZXShen2001Bogdanov, ZXShen2001DLFeng, DSDessau2001YDChuang, TXiang2003JChang}. It is interesting to ask whether the superconducting gap on the two Fermi surface sheets is the same or not because they have rather different doping levels.  Due to instrumental resolution limit, the previous ARPES measurements on Bi2212 all indicated that the two Fermi surface sheets have similar superconducting gap within the experimental uncertainty\cite{RFollath2002SVBorisenko, SUchida2013HAnzai}.

In this paper, we report high resolution laser-based ARPES measurements on the electronic structure and superconducting gap of Bi2212.  Using our latest generation of laser ARPES that can cover  two-dimensional momentum space at one time with super-high resolution, we have identified the bilayer splitting and measured the superconducting gap with unprecedented precision. In an overdoped Bi2212 with a T$_c$ of 75 K, we have observed two Fermi surface sheets which encompass rather different areas: the bonding Fermi surface sheet has a doping level of 0.14 which corresponds to a slightly underdoped case while the antibonding sheet has a doping level of 0.27 that corresponds to a heavily overdoped case. For the first time, we have revealed that the bonding and antibonding Fermi surface sheets have different superconducting gap. The momentum dependence of the superconducting gap for the antibonding Fermi surface sheet with 0.27 doping level follows the standard {\it d}-wave form while it deviates from the standard {\it d}-wave for the bonding Fermi surface with 0.14 doping level. This Fermi surface dependence of the superconducting gap is contrasted with that found in three-layer $\rm Bi_2Sr_2Ca_2Cu_3O_{10+\delta}$ (Bi2223)\cite{SUchida2010SIdetaPRL}. Our results provide important information for understanding the relationship between the Fermi surface topology and superconductivity, and also the layer-dependent superconductivity in cuprate superconductors.

\begin{figure*}[tbp]
\begin{center}
\includegraphics [width=0.6\columnwidth,angle=0]{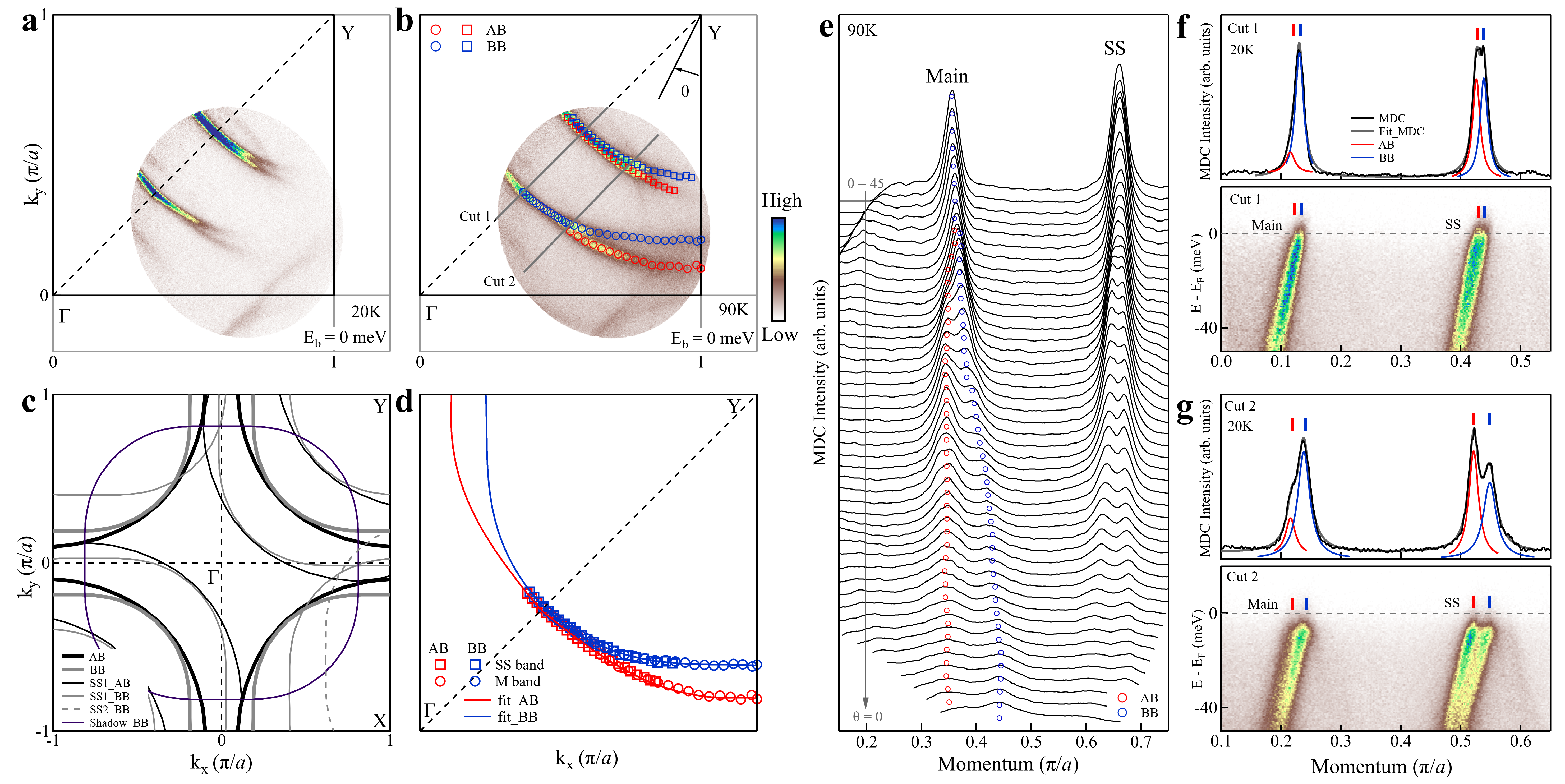}
\end{center}
\caption {\textbf{Fermi surface of  overdoped Bi2212 with a T$_c$ $\sim$ 75 K.}  (a) Fermi surface mapping  measured at  20 K.  (b) Fermi surface mapping measured at  90 K. The fitted Fermi momenta of the bonding Fermi surface sheet (BB) and antibonding sheet (AB) for both the main Fermi surface (Main) and the superstructure Fermi surface (SS) are marked.  (c) Schematic Fermi surface of Bi2212 containing the main Fermi surface: bonding sheet (thick grey lines) and antibonding sheet (thick black lines),  superstructure Fermi surface: first order of the bonding sheet (SS1BB, thin grey lines) and the antibonding sheet (SS1AB, thin black lines) and second order of the bonding sheet (SS2BB, dashed grey line),  and shadow Fermi surface of the bonding sheet (ShadowBB, violet line).  (d) Fermi surface obtained by combining the Fermi momenta from the main Fermi surface (circles) and superstructure Fermi surface (squares) from (b). The fitted bonding and antibonding Fermi surface sheets are plotted as thick blue and red lines, respectively. (e) Momentum distribution curves (MDCs) at the Fermi level (E$_F$) along various momentum cuts corresponding to the measurement at 90 K in (b).  The MDCs are offset along the vertical direction for clarity. The momentum cuts all pass through Y($\pi$,$\pi$) point, and are defined by the angle $\theta$ as indicated in (b) ($\theta$=45 corresponds to the nodal cut while $\theta$=0 corresponds to the ($\pi$,$\pi$)-($\pi$,0) antinodal cut). Two sets of peaks are observed corresponding to the main Fermi surface (Main) and superstructure Fermi surface (SS). The Main MDC peaks consist of two sub-peaks due to bilayer splitting: antibonding peak (red circles) and bonding peak (blue circles). The MDCs are fitted by two Lorentzians and the obtained peak positions are plotted in (b) as the Fermi momenta for the bonding and antibonding Fermi surface sheets.  The Fermi momenta for the superstructure Fermi surface in (b) are obtained in a similar manner.  (f-g) Band structure along the momentum cut 1 (f)  and cut 2 (g) measured at 20 K (lower panels). The location of the momentum cut 1 and cut 2, parallel to $\Gamma$-Y nodal direction,  is shown as grey lines in (b).  The upper panels shows the MDC at the Fermi level.  The main peak and the superstructure peak are fitted each by two Lorentzians that correspond to contributions from bonding band (blue) and antibonding band (red).
}
\label{Fig_FS}
\end{figure*}

High resolution angle-resolved photoemission measurements were performed using a new lab-based ARPES system equipped with  6.994 eV and 10.897 eV vacuum-ultra-violet (VUV) laser light source and the angle-resolved time-of-flight electron energy analyzer (ARToF)\cite{XJZhou2016CLWang, XJZhou2018LaserARPES}. One significant advantage of this ARToF-ARPES system is that it can cover two-dimensional momentum space simultaneously; with 10.897 eV laser as the light source, the detected momentum space can reach both the nodal and anti-nodal regions. In particular, it can cover nearly half a quadrant that includes both the nodal and antinodal regions for one measurement; then all the data are taken under the same experimental condition (Figs.~\ref{Fig_FS} a and 1b). The second advantage is its high instrumental resolution: the energy resolution is better than 1 meV and the angular resolution is 0.1$^\circ$ corresponding to 0.0023 $\rm {\AA}^{-1}$
momentum resolution for the photon energy of 10.897 eV at $\Gamma$ point. The third advantage is that, with the use of delay-line detector, it has much weaker non-linearity effect so that the measured signal is intrinsic to the sample.  The Fermi level is referenced by measuring on clean polycrystalline gold that is electrically connected to the sample  or referenced to the nodal direction of Bi2212 superconductors where the superconducting gap is known to be zero. The overdoped Bi2212 single crystals with a T$_c$ at 75 K were obtained by annealing the as-grown Bi2212 sample under high oxygen pressure\cite{XJZhou2016YXZhanganneal}.  The samples were cleaved \emph{in situ} at low temperature of 20 K and measured in ultrahigh vacuum with a base pressure better than 3$\times$10$^{-11}$ mbar.




Figure~\ref{Fig_FS} shows the Fermi surface of the overdoped Bi2212 (T$_c$=75 K) measured at 20 K in the superconducting state (Fig.~\ref{Fig_FS}a) and at 90 K in the normal state (Fig.~\ref{Fig_FS}b).  Using our laser-based ARToF-ARPES system with a photon energy of 10.897 eV,  more than half of a quadrant of the first Brillouin zone can be covered simultaneously which encompasses both the nodal and antinodal regions of Bi2212. In Bi2212, since there are two CuO$_2$ planes in one structural unit (half a unit cell along {\it c} direction) separated by calcium, the interaction between the two CuO$_2$ planes gives rise to bilayer splitting that results in two Fermi surface sheets: bonding sheet (BB) and antibonding (AB) sheet, as depicted in  Fig.~\ref{Fig_FS}c as thick lines\cite{FPaulsen1995OKAndersenNew, ABansil1999PRL, ZXShen2001Bogdanov, ZXShen2001DLFeng, DSDessau2004YDChuang}. This bilayer splitting is more pronounced in overdoped Bi2212 samples\cite{DSDessau2004YDChuang}. In addition, because Bi2212 has a superstructure modulation along b$^*$ ($\Gamma$-Y) direction\cite{AWSleight1988YAGao, OEibl1991, GKostorz1994HHeinrich, BGHyde1988RLWithers}, it produces superstructure Fermi surface with different orders, n,  as shown in Fig.~\ref{Fig_FS}c by thin lines  (n=1, the first order) or dashed line (n=2, the second order), by shifting the original main Fermi surface with vectors $\pm$nQ where Q refers to the superstructure modulation vector\cite{Aebi1994, Osterwalder1995, GPBrivio1996HDing, KKadowaki2000HMFretwell}. Moreover, shadow Fermi surface is also present which represents a replica of the main Fermi surface shifted by Q=($\pi$,$\pi$), as shown by the violet line in Fig.~\ref{Fig_FS}c\cite{Aebi1994}. The observed features in Fig.~\ref{Fig_FS}a and Fig.~\ref{Fig_FS}b can be well accounted for by the main Fermi surface with bilayer splitting, the corresponding superstructure Fermi surface, and the shadow Fermi surface, as shown in Fig.~\ref{Fig_FS}c.  Particularly prominent features are the main Fermi surface (marked as Main in Fig.~\ref{Fig_FS}b) and one branch of the superstructure Fermi surface (marked as SS in Fig.~\ref{Fig_FS}b).  Fig.~\ref{Fig_FS}e shows the momentum distribution curves (MDCs) at the Fermi level along the momentum cuts that all cross the Y($\pi$,$\pi$) point, covering the measured area, and defined by the angle $\theta$ as marked in Fig.~\ref{Fig_FS}b. Two groups of peaks are observed that correspond to the main Fermi surface (Main) and superstructure Fermi surface (SS). The main peak in Fig.~\ref{Fig_FS}e consists of two sub-peaks that correspond to the antibonding band (red circles in Fig.~\ref{Fig_FS}e) and bonding band (blue circles in Fig.~\ref{Fig_FS}e).  The main peaks are fitted by two Lorentzians and the fitted peak positions give the Fermi momenta of the antibonding band (red circles in Fig.~\ref{Fig_FS}b) and bonding band (blue circles in Fig.~\ref{Fig_FS}b). The superstructure Fermi surface can be analyzed in the similar manner, and the measured Fermi momenta of the antibonding band (red squares in Fig.~\ref{Fig_FS}b) and bonding band (blue squares in Fig.~\ref{Fig_FS}b) are plotted in Fig.~\ref{Fig_FS}b. Since the superstructure Fermi surface is a replica of the main Fermi surface shifted by a superstructure modulation vector, Q=(0.218,0.218)$\pi$/a for the present sample, the two sets of Fermi surface sheets obtained from the main and superstructure bands are equivalent.  They are combined in Fig.~\ref{Fig_FS}d to give an overall Fermi surface of our measured sample: antibonding Fermi surface sheet (red line in Fig.~\ref{Fig_FS}d) and bonding Fermi surface sheet (blue line in Fig.~\ref{Fig_FS}d).

As shown above in Fig.~\ref{Fig_FS}, the bilayer splitting is clearly observed in the Bi2212 sample by our laser-ARPES measurement. In particular, the bilayer splitting along the nodal direction can be clearly resolved in the present measurement.  Taking advantage of the photoemission matrix element effects by using different photon energies, such a bilayer splitting along the nodal direction was resolved before for the main Fermi surface
\cite{Lindroos2004PRB, YAndo2004AAKordyuk, STurchini2004SVBorisenko, DSDessau2007HIwasawa, SUchida2013HAnzai}.
As seen in Fig.~\ref{Fig_FS}a and Fig.~\ref{Fig_FS}b, the bonding band and antibonding band show different matrix element effects in the main Fermi surface and in the superstructure Fermi surface. The bilayer splitting around the nodal region is more clearly resolved in the superstructure Fermi surface other than the main Fermi surface. From the band structure and corresponding MDCs in Fig.~\ref{Fig_FS}f and 1g measured along two momentum cuts around the nodal region, the signal of the main Fermi surface is dominated by the bonding band; the antibonding band is suppressed and contributes only a small fraction of the signal.  For the superstructure Fermi surface, the signals of the bonding band and the antibonding band are comparable, as seen from the MDCs in Fig.~\ref{Fig_FS}f and Fig.~\ref{Fig_FS}g. The different matrix element effects for the main Fermi surface and the superstructure Fermi surface explains why the bilayer splitting around the nodal region is more clearly resolved in the superstructure Fermi surface, not in the main Fermi surface. This peculiar matrix element effects also makes it possible for us to resolve the bilayer splitting in Bi2212 around the entire Fermi surface.  Along the nodal direction, the bilayer splitting we observed between the bonding and antibonding bands is 0.012$\pi$/a for the overdoped Bi2212 sample with a T$_c$=75 K we have measured, obtained by the separation of the two fitted Lorentzian peaks for the superstructure MDC peak in Fig.~\ref{Fig_FS}f, which is consistent with previous measurements\cite{STurchini2004SVBorisenko, YAndo2004AAKordyuk, DSDessau2007HIwasawa}.

Our high resolution laser-ARPES measurements provide a precise Fermi surface determination of Bi2212 by measuring a large momentum space simultaneously under the same condition that covers both the nodal and antinodal regions, resolving the bonding and antibonding bands along the entire Fermi surface,  combing the results from both the main band and the superstructure band, and measuring both the normal state and the superconducting state. The obtained bonding Fermi surface sheet and antibonding Fermi surface sheet are shown in Fig.~\ref{Fig_FS}d. When neglecting the weak k$_z$ effect of the Fermi surface in Bi2212\cite{ABansil2005RSMarkiewicz},  and from the Luttinger volume of the Fermi surface, the antibonding Fermi surface sheet (red line in Fig.~\ref{Fig_FS}d) corresponds to a doping level of 0.14 while the bonding Fermi surface sheet (blue line in Fig.~\ref{Fig_FS}d) corresponds to a doping level of 0.27.  Considering that the optimal doping of Bi2212 lies near {\it p}$\sim$0.16\cite{ZXShen2012IMVIshik},  the bonding Fermi surface sheet  ({\it p}$\sim$0.14) can be taken as slightly underdoped one while the antibonding Fermi surface sheet ({\it p}$\sim$0.27) can be taken as heavily overdoped one.



\begin{figure*}[tbp]
\begin{center}
\includegraphics [width=1.0\columnwidth,angle=0]{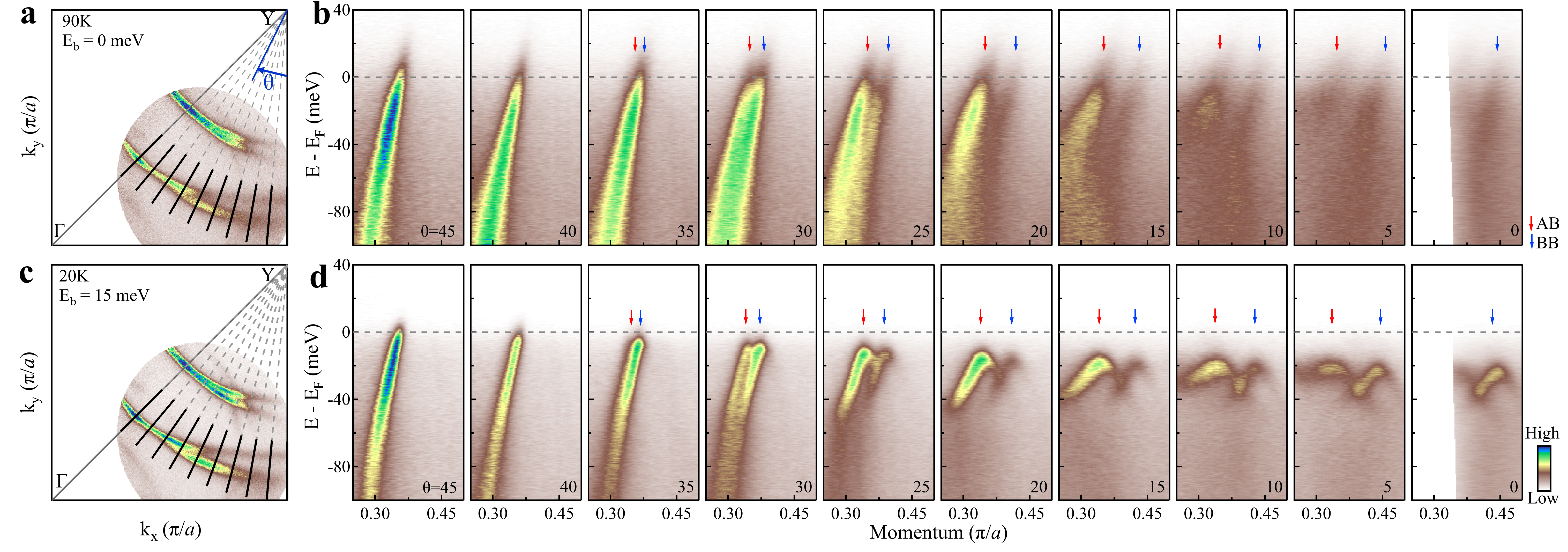}
\end{center}
\caption {\textbf{Band structure of Bi2212 in normal and superconducting states.} (a) Fermi surface mapping at 90 K in the normal state.  (b) Band structure of the main Fermi surface at 90 K along different momentum cuts.  The location of the momentum cuts is shown in (a) as black lines. These lines all point to the Y ($\pi$,$\pi$)  and are labeled by the angle $\theta$ defined in (a).  (c,d) are the same as (a,b) but measured at 20 K in the superconducting state. The bonding band (BB) and antibonding band (AB)  are marked by blue arrows and red arrows, respectively, in (b) and (d).
}
\label{Fig_bandstructure}
\end{figure*}

Figure \ref{Fig_bandstructure} shows the Bi2212 band structure of the main Fermi surface measured in both the normal state and superconducting state.  The two-dimensional momentum coverage makes it possible to align the sample orientation precisely that can avoid the sample orientation change during the process of varying the sample temperature.  It also makes it possible to take the band structure along any momentum cut in the covered momentum space. To facilitate direct comparison between the bonding and antibonding bands, we take the momentum cuts as shown in Fig.~\ref{Fig_bandstructure}a by the black lines. These momentum cuts all pass through the Y($\pi$,$\pi$) point and are defined by the angle $\theta$ as shown in Fig.~\ref{Fig_bandstructure}a. Fig.~\ref{Fig_bandstructure}b shows the band structure along these momentum cuts measured at 90 K in the normal state. The corresponding band structure measured at 20 K in the superconducting state is shown in Fig.~\ref{Fig_bandstructure}d, with the same location of the momentum cuts marked in Fig.~\ref{Fig_bandstructure}c. Both the bonding band and the antibonding band are marked in Fig.~\ref{Fig_bandstructure}b and Fig.~\ref{Fig_bandstructure}d; they split larger and larger when the momentum cuts approach the antinodal region.  The band structure experiences a significant change upon entering the superconducting state, especially near the antinodal region. In the superconducting state, signatures of Bogoliubov bands for both the bonding and antibonding bands, caused by superconducting gap opening,  are clearly seen and there appears interaction between the bonding and antibonding bands (Fig.~\ref{Fig_bandstructure}d).  The dramatic changes in the superconducting state are due to a combined effect of both the superconducting gap opening and the mode coupling\cite{Damascelli2003ZXShen, XJZhouReview, NNagaosa2005TCuk, XJZhou2013JFHe}. We also note that the relative intensity of the bonding and antibonding bands along the main Fermi surface shows a peculiar momentum dependence (Fig. 2). Near the nodal region ($\theta=45\sim30$), the signal of the bonding band dominates. In the intermediate region ($\theta=25\sim10$), the antibonding band becomes dominant. Near the antinodal region ($\theta=5\sim0$), the intensity of the two bands is comparable. The intensity variation with momentum is similar for both measurements above T$_c$ (Figs. 2a and 2b) and below T$_c$ (Figs. 2c and 2d). Therefore, such an intensity change with momentum for the bonding and antibonding bands is more likely related to the photoemission matrix element effects\cite{Damascelli2003ZXShen, TXiang2003JChang}. In the following, we will mainly focus on the superconducting gap of the bonding and antibonding Fermi surface sheets in Bi2212.

\begin{figure*}[tbp]
\begin{center}
\includegraphics [width=1.0\columnwidth,angle=0]{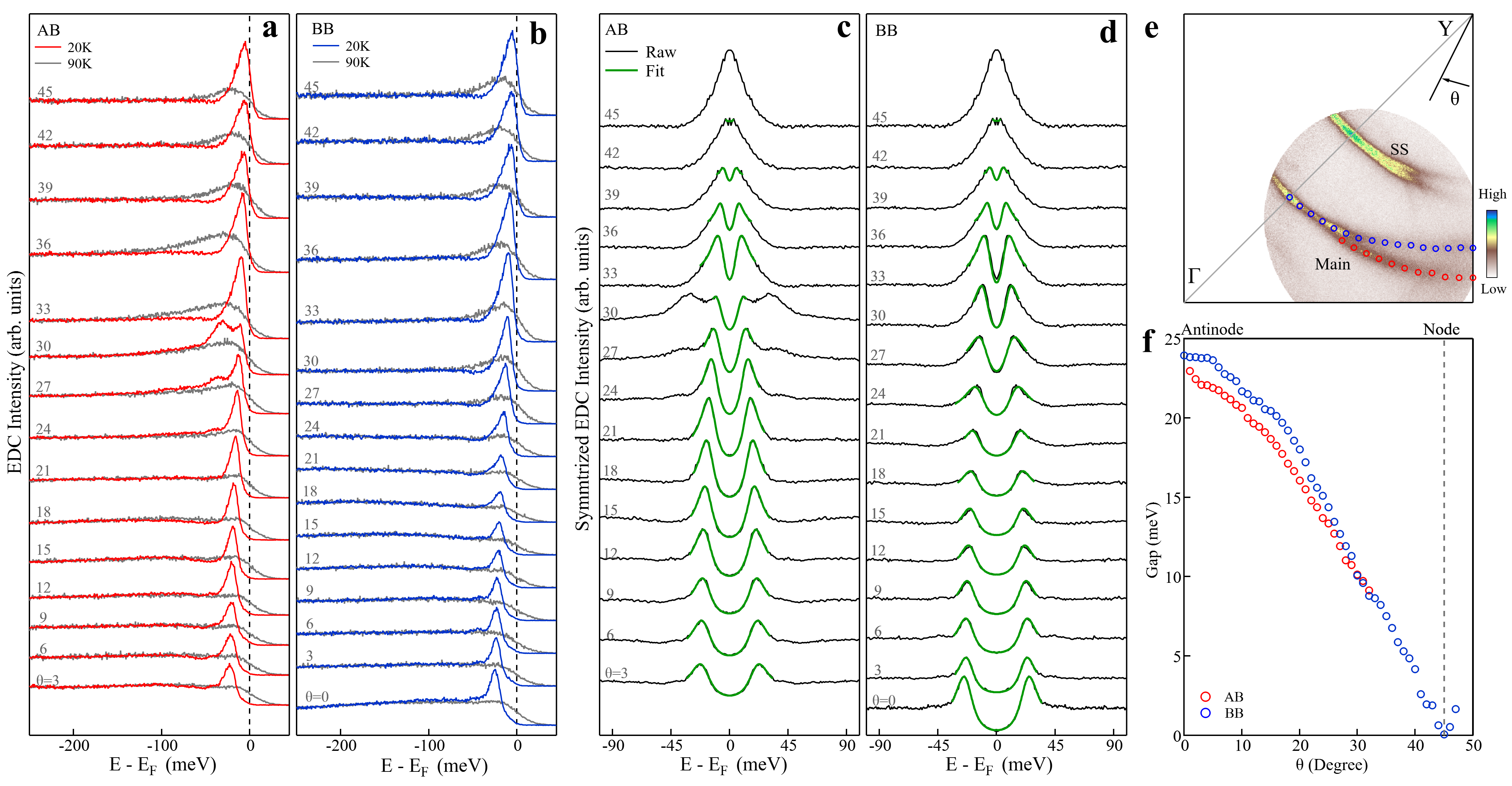}
\end{center}
\caption {\textbf{Photoemission spectra along the two main Fermi surface sheets and determination of the superconducting gap.} (a,b) Photoemission spectra (energy distribution curves, EDCs) along the antibonding Fermi surface sheet (a) and bonding Fermi surface sheet (b). Both the normal state EDCs (90 K, grey lines in (a) and (b)) and EDCs in the superconducting state (20 K, red lines in (a) and blue line in (b)) are plotted.  The corresponding symmetrized EDCs of the 20 K data for the antibonding band and bonding band are shown in (c) and (d), respectively. Green lines are fitted curves from the phenomenological gap formula\cite{NormanPRB1998}.  The location of the Fermi momenta are marked in (e) on the measured Main Fermi surface where red circles represents the momenta along the antibonding Fermi surface while the blue circles represents momenta along the bonding Fermi surface. The momentum points are labeled by the angle $\theta$ in (a-d) that is defined in (e).  (f) The superconducting gap as a function of $\theta$ for the antibonding band (red circles) and bonding band (blue circles) obtained by fitting the symmetrized EDCs in (c) and (d), respectively.
}
\label{Fig_EDC1}
\end{figure*}

In order to extract the superconducting gap, the photoemission spectra (energy distribution curves, EDCs) along the antibonding Fermi surface sheet and bonding Fermi surface sheet of the Main Fermi surface are picked up and shown in Figs.~\ref{Fig_EDC1}a and 3b, respectively. For comparison, the EDCs measured in the normal state (90 K) and superconducting state (20 K) are both plotted. The corresponding symmetrized EDCs measured at 20 K from Figs.~\ref{Fig_EDC1}a and 3b are shown in Figs.~\ref{Fig_EDC1}c and 3d, respectively. The location of the Fermi momenta on the main Fermi surface are marked in Fig.~\ref{Fig_EDC1}e that are defined by the angle $\theta$. Sharp superconducting coherence peaks develop in the superconducting state, from the broad EDCs in the normal state,  along the bonding and antibonding Fermi surface sheets (Figs.~\ref{Fig_EDC1}a and 3b).  The symmetrized EDCs are fitted by a phenomenological gap formula\cite{NormanPRB1998},  taking the self-energy $\Sigma(k,\omega)=-i\Gamma_1 + \Delta^2/[(\omega+i0^{+})+\epsilon(k)]$. This is a standard procedure to extract superconducting gap in superconductors where $\Gamma_1$ is a single-particle scattering rate, $\Delta$ is the superconducting gap and $\epsilon(k)$ is the band dispersion that is 0 at the Fermi level\cite{NormanPRB1998}. The extracted superconducting gap for the bonding and antibonding Fermi surface along the main Fermi surface, as a function of the angle $\theta$,  is shown in Fig.~\ref{Fig_EDC1}f. Near the nodal region ($\theta = 30^\circ \sim 45^\circ$), bilayer splitting is not well-resolved along the main Fermi surface because the signal is dominated by the bonding band, as shown in Fig.~\ref{Fig_FS}. Therefore, the superconducting gap obtained in this nodal region mainly represents the gap of the bonding Fermi surface sheet (Fig.~\ref{Fig_EDC1}f). When the bilayer splitting becomes resolvable between $\theta=0^\circ$ and 30$^\circ$, the superconducting gap of the bonding and antibonding Fermi surface sheets can be precisely determined, and is different near the antinodal region (Fig.~\ref{Fig_EDC1}f).

\begin{figure*}[tbp]
\begin{center}
\includegraphics [width=0.7\columnwidth,angle=0]{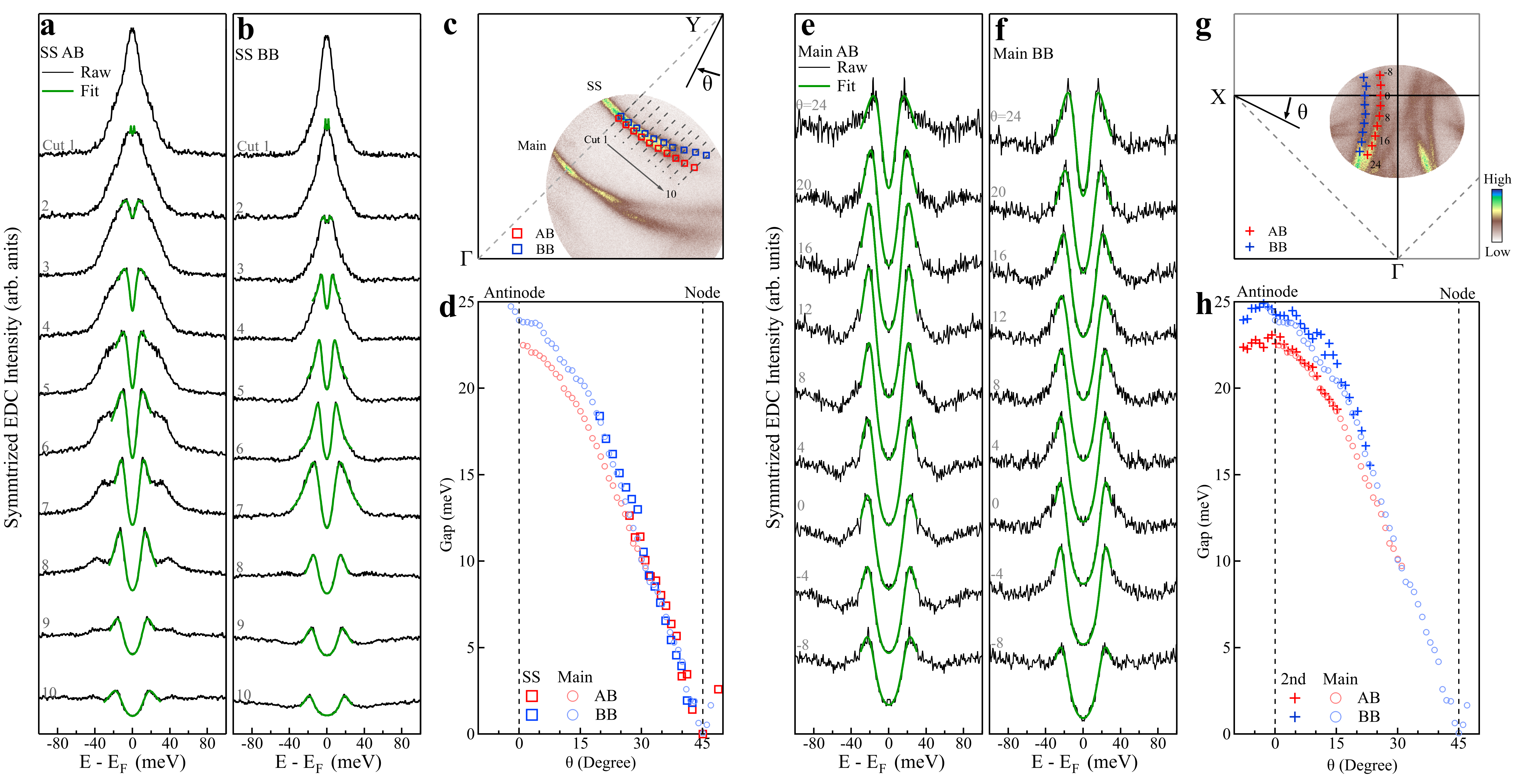}
\end{center}
\caption {\textbf{Determination of the superconducting gap for the bonding and antibonding Fermi surface sheets near nodal and antinodal regions.}  (a,b)  Symmetrized EDCs measured at 20 K along the antibonding Fermi surface sheet and bonding sheet of the superstructure Fermi surface (SS) in the Fermi surface mapping in (c). The green lines in (a) and (b) represent the fitting curves by using the phenomenological gap formula\cite{NormanPRB1998}. The location of the Fermi momentum points are marked as red and blue circles for the antibonding and bonding Fermi surface sheets, respectively, in (c) along ten different momentum cuts. These momentum cuts are all parallel to the $\Gamma$(0,0)-Y($\pi$,$\pi$) nodal direction.  (d) Superconducting gap as a function of the $\theta$ angle for the antibonding (red rectangles) and bonding (blue rectangles) Fermi surface sheets in the superstructure Fermi surface, determined from fitting the symmetrized EDCs as shown in (a) and (b). For comparison, the gap determined from the main Fermi surface in Fig.~\ref{Fig_EDC1}f is also included.  Also for a direct comparison, we use the same $\theta$ angle definition as in Figs.~\ref{Fig_FS}-\ref{Fig_EDC1} by shifting the superstructure Fermi surface sheets to the position of the main Fermi surface along the nodal direction with a superstructure wave vector (-0.218,-0.218)$\pi$/a. (e,f)  Symmetrized EDCs measured at 20 K along the antibonding (e) and bonding (f) Fermi surface sheets of the main Fermi surface from another Fermi surface mapping around the (0,$\pi$) antinodal region,  as shown in (g). The green lines in (e) and (f) represent the fitting curves by using the phenomenological gap formula\cite{NormanPRB1998}. The location of the Fermi momentum points are marked as red and blue crosses for the antibonding and bonding Fermi surface sheets, respectively, in (g).  (h) Superconducting gap as a function of the $\theta$ angle for the antibonding (red crosses) and bonding (blue crosses) Fermi surface sheets determined from fitting the symmetrized EDCs as shown in (e) and (f). For comparison, the gap determined from the main Fermi surface in Fig.~\ref{Fig_EDC1}f is also shown.  Also for a direct comparison, we use the same $\theta$ angle definition in (g) as in Figs.~\ref{Fig_FS}-\ref{Fig_EDC1}, i.e., $\theta=45^\circ$ corresponds to the nodal cut and $\theta$=0 corresponds to the antinodal cut.
 }
\label{Fig_EDC2}
\end{figure*}

The clear observation of the bilayer splitting near the nodal region along the superstructure Fermi surface (Fig.~\ref{Fig_FS}) provides a chance to extract the superconducting gap on the bonding and antibonding Fermi surface sheets near the nodal region that is not accessible from the main Fermi surface (Fig.~\ref{Fig_EDC1}). Figs.~\ref{Fig_EDC2}a and 4b show the symmetrized EDCs measured at 20 K along the antibonding Fermi surface sheet and bonding sheet of the superstructure Fermi surface, respectively. The location of the Fermi momenta is shown in Fig.~\ref{Fig_EDC2}c which are along ten parallel momentum cuts. The symmetrized EDCs are fitted by the phenomenological gap formula in Figs.~\ref{Fig_EDC2}a and 4b, and the obtained superconducting gap for the antibonding Fermi surface sheet and bonding sheet of the superstructure Fermi surface is shown in Fig.~\ref{Fig_EDC2}d. Here the $\theta$ angle follows the same definition as in Fig.~\ref{Fig_EDC1}e by shifting the superstructure Fermi surface to the position of the main Fermi surface because the former is a replica of the latter. It can be seen from Fig.~\ref{Fig_EDC2}d that the superconducting gap is nearly the same within our experimental uncertainty near the nodal region for the bonding and antibonding Fermi surface sheets. They are also consistent with the superconducting gap measured from the main Fermi surface.

In order to further examine the superconducting gap near the antinodal region, we took another measurement focusing on the (0,$\pi$) antinodal region, as shown in Fig.~\ref{Fig_EDC2}g. Bonding and antibonding Fermi surface sheets are clearly resolved on both left and right sides of the $\Gamma$(0,0)-(0,$\pi$) line (Fig.~\ref{Fig_EDC2}g). The symmetrized EDCs along the antibonding Fermi surface and bonding Fermi surface in this antinodal area are shown in Fig.~\ref{Fig_EDC2}e and Fig.~\ref{Fig_EDC2}f, respectively. The location of the Fermi momenta is marked in Fig.~\ref{Fig_EDC2}g. The symmetrized EDCs in Figs.~\ref{Fig_EDC2}e and 4f are fitted by the phenomenological gap formula and the obtained superconducting gap is plotted in Fig.~\ref{Fig_EDC2}h.  Here the angle $\theta$ is defined in accordance with the one in Fig.~\ref{Fig_EDC1}e that it is 45$^\circ$ along the nodal direction while it is 0$^\circ$ along the antinodal direction. The data are consistent with those obtained from Fig.~\ref{Fig_EDC1} and the bonding and antibonding Fermi surface sheets exhibit different superconducting gap near the antinodal region (Fig.~\ref{Fig_EDC2}h).

\begin{figure*}[tbp]
\begin{center}
\includegraphics [width=1.0\columnwidth,angle=0]{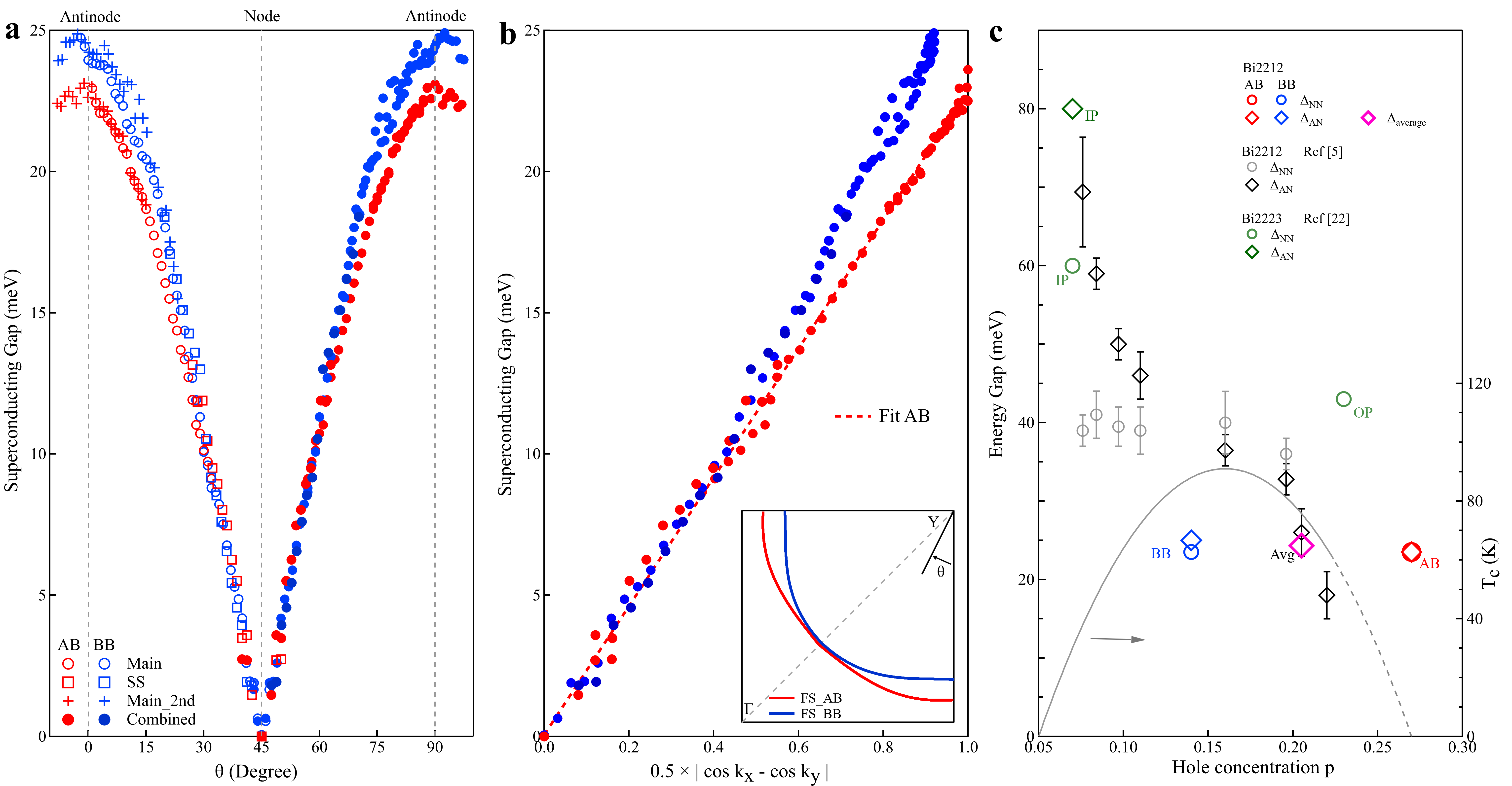}
\end{center}
\caption {\textbf{Distinct superconducting gap of the bonding and antibonding Fermi surface sheets in Bi2212.} (a) Superconducting gap as a function of the $\theta$ angle defined in the inset of (b) for the antibonding Fermi surface sheet (red symbols) and bonding Fermi surface sheet (blue symbols). The data summarize all the gap measurements in Fig.~\ref{Fig_EDC1}f, Fig.~\ref{Fig_EDC2}d and Fig.~\ref{Fig_EDC2}h, and are symmetrized with respect to the nodal direction ($\theta$=45).  (b) Measured superconducting gap of the antibonding Fermi surface sheet (red circles) and bonding Fermi surface sheet (blue circles) as a function of 0.5$\times$$|$cosk$_x$-cosk$_y$$|$. The measured Fermi surface is shown in the bottom-right inset. The dashed line represents a linear fit to the superconducting gap of the antibonding Fermi surface (red circles).   (c) Schematic phase diagram of Bi2212\cite{ZXShen2012IMVIshik}. The grey line represents variation of the superconducting transition temperature (T$_c$) with the hole concentration.  The superconducting gap of Bi2212 with different doping levels measured from ARPES is plotted: the antinodal gap ($\Delta_{AN}$, black diamonds) that is directly measured from the EDCs at the antinodal region and the so-called nodal gap ($\Delta_{NN}$, grey circles) that is obtained by extrapolating the gap near the nodal region to the antinodal region\cite{ZXShen2012IMVIshik}.  For comparison, the superconducting gap of Bi2223 is also included which includes the nodal gap of the outer CuO$_2$ plane and the inner CuO$_2$ plane ($\Delta_{NN}$, green circles) and antinodal gap of the inner CuO$_2$ plane ($\Delta_{AN}$, green diamond)\cite{SUchida2010SIdetaPRL}. For the present Bi2212 measurements,  the nodal gap of the antibonding (red circle)  and bonding (blue circle) Fermi surface,  the antinodal gap of the antibonding (red diamond) and bonding (blue diamond) Fermi surface, and the antinodal gap for the averaged doping level of bonding and antibonding Fermi surface (pink diamond) are plotted.
}
\label{Fig_gap}
\end{figure*}

Figure \ref{Fig_gap}a summarizes the superconducting gap measurements from the main Fermi surface (Fig.~\ref{Fig_EDC1}g), the superstructure Fermi surface (Fig.~\ref{Fig_EDC2}d) and the main Fermi surface near the antinodal region (Fig.~\ref{Fig_EDC2}h). For the first time, it is revealed that the bonding and antibonding Fermi surface sheets have a different superconducting gap in Bi2212; the maximum gap size difference near the antinodal region is $\sim$2 meV. Previous ARPES measurements did not resolve such a difference\cite{RFollath2002SVBorisenko, SUchida2013HAnzai}. Super-high instrumental resolution and simultaneous two-dimensional momentum coverage of our laser-based ARToF-ARPES system play key roles in achieving such a conclusion. The superconducting gap is also plotted as a function of the standard {\it d}-wave form: 0.5$\times$$|$cosk$_x$-cosk$_y$$|$ in Fig.~\ref{Fig_gap}b. The superconducting gap for the antibonding Fermi surface (red circles in Fig.~\ref{Fig_gap}b) follows basically a straight line and obeys the standard {\it d}-wave form. On the other hand, the gap for the bonding Fermi surface follows a linear line near the nodal region but deviates from the linear line in the antinodal region. Such a deviation may be interpreted as the signature of pseudogap formation\cite{ZXShen2012IMVIshik} or high harmonics of the superconducting gap\cite{KKadowaki1999JMesot} which becomes more obvious in the underdoped region. In this case, two kinds of gaps can be defined: one is the antinodal gap measured directly near the antinodal region, $\Delta_{AN}$, the other is the so-called nodal gap that is obtained by fitting the superconducting gap near the nodal region with a linear line and extrapolate to the antinodal region, $\Delta_{NN}$\cite{ZXShen2012IMVIshik}. According to the definition, the antinodal gap of our Bi2212 sample is $\sim$25 meV for the bonding Fermi surface (blue diamond in Fig.~\ref{Fig_gap}c) and $\sim$23 meV for the antibonding Fermi surface (red diamond in Fig.~\ref{Fig_gap}c).  The nodal gap for the bonding Fermi surface (blue circle in Fig.~\ref{Fig_gap}c) and antibonding Fermi surface (red circle in Fig.~\ref{Fig_gap}c) is nearly the same ($\sim$23 meV).

It is known that the measured superconducting gap of Bi2212 exhibits an unusual behavior: the antinodal gap ($\Delta_{AN}$) increases with decreasing doping while the nodal gap  ($\Delta_{NN}$) is not sensitive to doping in the optimally and underedoped samples (Fig.~\ref{Fig_gap}c)\cite{ZXShen2012IMVIshik}. However, these data are obtained by mainly considering an overall average doping level in Bi2212. As shown in Fig.~\ref{Fig_gap}c, if we plot the average gap size as a function of the average doping level for our sample, it is consistent with the previous results. However,  since Bi2212 contains two Fermi surface sheets with different doping levels, it is more reasonable to consider the superconducting gap on each individual Fermi surface sheet.  As shown in Fig.~\ref{Fig_gap}c, if we plot the superconducting gap as a function of doping for the two separate Fermi surface sheets, the relation obviously deviates from the usual overall behavior. In this case, the bonding Fermi surface with a doping of 0.14 that is slightly underdoped has a superconducting gap that is significantly smaller than the Bi2212 with an overall average doping level of 0.14.  The antibonding Fermi surface with a doping of 0.27 that is heavily overdoped still has a significantly large superconducting gap while for the Bi2212 sample with an overall average doping of 0.27, it has nearly zero gap and becomes non-superconducting. These indicate that the superconducting gap associated with a specific Fermi surface in Bi2212 is not solely determined by the Fermi surface topology or its doping level. The superconductivity is an collective effect that is associated with the overall superconducting unit and the interaction between the two CuO$_2$ planes in one structural unit.

It is also instructive to compare the relationship between the superconducting gap and the Fermi surface topology between Bi2223 and Bi2212.  In Bi2223 with triple CuO$_2$ planes in one structural unit, one would expect three Fermi surface sheets (bonding, antibonding and nonbonding) through the interaction between the three CuO$_2$ planes. Instead, two distinct Fermi surface sheets  are observed in the optimally-doped Bi2223 with a T$_c$=110 K\cite{SUchida2010SIdeta}. These two Fermi surface sheets are attributed to the inner CuO$_2$ plane with a lower doping of 0.07 and the two outer CuO$_2$ planes with a high doping of 0.23, giving rise to an overall average doping level of 0.18.  It was also found that the superconducting gap for the outer Fermi surface (0.23 doping level) follows a standard {\it d}-wave from, while it deviates from the standard {\it d}-wave form for the inner Fermi surface\cite{SUchida2010SIdeta}.  The Fermi surface topology of our present Bi2212 sample shows a strong resemblance to that of Bi2223. In both cases, two hole-like Fermi surface sheets are observed, one is underdoped and the other is heavily overdoped.  The superconducting gap of Bi2212 and Bi2223  is also similar in that the gap of the overdoped Fermi surface sheet follows a standard {\it d}-wave form while the underdoped Fermi surface deviates.   The most obvious difference between Bi2212 and Bi2223 lies in the absolute gap size and the maximum gap difference between the two Fermi surface sheets.  In Bi2223,  the absolute value of the gap is much larger (Fig.~\ref{Fig_gap}c), the maximum gap size difference between the inner and outer Fermi surface sheets can reach up to $\sim$38 meV (Fig.~\ref{Fig_gap}c)\cite{SUchida2010SIdeta,TKondo2017SKunisada}.  In contrast,  for Bi2212 in our present case, the maximum gap difference between the bonding and antibonding Fermi surface sheets is only $\sim$2 meV (Fig.~\ref{Fig_gap}c).  This gap difference appears rather small considering the dramatic doping level difference between the two Fermi surface sheets. This disparity of the gap size difference between the two Fermi surface sheets may be attributed to the different origin of the two Fermi surface in Bi2212 and Bi2223.  In Bi2223, the doping level of the inner CuO$_2$ plane is much smaller than that in the outer two CuO$_2$ planes, giving rise to two distinct Fermi surface sheets corresponding to the two kinds of CuO$_2$ planes with rather different doping levels\cite{SUchida2010SIdeta}.  In Bi2212, the two CuO$_2$ planes in one unit is  structurally identical;  the two Fermi surface sheets originate from the hybridization of electronic structures between the two CuO$_2$ planes. It was expected that the two Fermi surface sheets in Bi2212 would give rise to similar superconducting gap for the bonding and antibonding Fermi surface sheets, as previous measurements indicated\cite{RFollath2002SVBorisenko, SUchida2013HAnzai}.  With the distinct superconducting gap revealed along the two Fermi surface sheets in Bi2212 (Fig.~\ref{Fig_gap}a),  it asks for further theoretical investigations to understand the gap difference when considering that the two Fermi surface sheets come from two structurally identical CuO$_2$ planes in real space, and  from fully hybridized electronic structure between the two CuO$_2$ planes in the reciprocal space.  The comparison between Bi2212 and Bi2223  again indicates that the superconducting gap of a specific Fermi surface sheet is not solely determined by its topology and the related doping level. The collective interaction between these Fermi surface sheets is important to dictate superconductivity. Compared with the single-layer Bi$_2$Sr$_2$CuO$_{6+\delta}$ (Bi2201) superconductor that has a single Fermi surface sheet with a lower maximum T$_c$ ($\sim$32 K) and smaller superconducting gap ($\sim$15 meV)\cite{XJZhou2009JQMeng},  the multiple Fermi surface sheets present in bilayer Bi2212 and trilayer Bi2223, and their interactions may provide extra ingredients to boost superconductivity in the cuprate superconductors\cite{SUchida2010SIdeta}.

In summary,  by taking high resolution laser-ARPES measurements on Bi2212, we have revealed for the first time the superconducting gap difference between the bonding and antibonding Fermi surface sheets that are caused by bilayer splitting in Bi2212. The superconducting gap along the antibonding Fermi surface with a doping of 0.27 follows the standard {\it d}-wave form, while the gap along the bonding Fermi surface with a doping of 0.14 deviates from the standard {\it d}-wave form. The maximum gap difference between the two Fermi surface sheets is $\sim$ 2 meV near the antinodal region.  These results provide important information to study the relationship between the Fermi surface topology, its associated superconducting gap and superconductivity, and the layer-dependence of superconductivity in high temperature cuprate superconductors.

\vspace{3mm}

\noindent {\bf Acknowledgement}  We thank financial support from the National Natural Science Foundation of China  (Grant No. 11888101),  the National Key Research and Development Program of China (Grant No. 2016YFA0300300 and 2017YFA0302900), the Strategic Priority Research Program (B) of the Chinese Academy of Sciences (XDB25000000),  the Youth Innovation Promotion Association of CAS (Grant No.2017013), and the Research Program of Beijing Academy of Quantum Information Sciences (Grant No. Y18G06).  The work at Brookhaven was supported by the Office of Basic Energy Sciences, U.S. Department of Energy (DOE) under Contract No. de-sc0012704.

\vspace{3mm}

\vspace{3mm}


\bibliographystyle{unsrt}


\renewcommand\figurename{Fig.}

\newpage

\end{document}